\documentclass[aps,prl,amsmath,amssymb,tightenlines,epsfig,floatfix, twocolumn]{revtex4-1}
\usepackage{graphicx}
\usepackage{dcolumn}	
\usepackage{bm}			
\usepackage{amsfonts}
\usepackage{xspace}
\usepackage{color}
\usepackage{epstopdf}
\usepackage{multirow}

\begin{document}

\newcommand{\psihat}{\ensuremath{\hat{\psi}}\xspace}
\newcommand{\psihatd}{\ensuremath{\hat{\psi}^{\dagger}}\xspace}
\newcommand{\ahat}{\ensuremath{\hat{a}}\xspace}
\newcommand{\Ham}{\ensuremath{\mathcal{H}}\xspace}
\newcommand{\ahatd}{\ensuremath{\hat{a}^{\dagger}}\xspace}
\newcommand{\bhat}{\ensuremath{\hat{b}}\xspace}
\newcommand{\bhatd}{\ensuremath{\hat{b}^{\dagger}}\xspace}
\newcommand{\boldr}{\ensuremath{\mathbf{r}}\xspace}
\newcommand{\dr}{\ensuremath{\,d^3\mathbf{r}}\xspace}
\newcommand{\tr}{\ensuremath{\,\mathrm{Tr}}\xspace}
\newcommand{\dk}{\ensuremath{\,d^3\mathbf{k}}\xspace}
\newcommand{\etal}{\emph{et al.\/}\xspace}
\newcommand{\ie}{i.e.}
\newcommand{\eq}[1]{Eq.~(\ref{#1})\xspace}
\newcommand{\fig}[1]{Fig.~\ref{#1}\xspace}
\newcommand{\abs}[1]{\left| #1 \right|}
\newcommand{\proj}[2]{\left| #1 \rangle\langle #2\right| \xspace}
\newcommand{\Qhat}{\ensuremath{\hat{Q}}\xspace}
\newcommand{\Qhatd}{\ensuremath{\hat{Q}^\dag}\xspace}
\newcommand{\phihatd}{\ensuremath{\hat{\phi}^{\dagger}}\xspace}
\newcommand{\phihat}{\ensuremath{\hat{\phi}}\xspace}
\newcommand{\boldk}{\ensuremath{\mathbf{k}}\xspace}
\newcommand{\boldp}{\ensuremath{\mathbf{p}}\xspace}
\newcommand{\boldsigma}{\ensuremath{\boldsymbol\sigma}\xspace}
\newcommand{\boldalpha}{\ensuremath{\boldsymbol\alpha}\xspace}
\newcommand{\grad}{\ensuremath{\boldsymbol\nabla}\xspace}
\newcommand{\parti}[2]{\frac{ \partial #1}{\partial #2} \xspace}
 \newcommand{\vs}[1]{\ensuremath{\boldsymbol{#1}}\xspace}
\renewcommand{\v}[1]{\ensuremath{\mathbf{#1}}\xspace}
\newcommand{\Psihat}{\ensuremath{\hat{\Psi}}\xspace}
\newcommand{\Psihatd}{\ensuremath{\hat{\Psi}^{\dagger}}\xspace}
\newcommand{\Vhatd}{\ensuremath{\hat{V}^{\dagger}}\xspace}
\newcommand{\Xhat}{\ensuremath{\hat{X}}\xspace}
\newcommand{\Xhatd}{\ensuremath{\hat{X}^{\dag}}\xspace}
\newcommand{\Yhat}{\ensuremath{\hat{Y}}\xspace}
\newcommand{\Jhat}{\ensuremath{\hat{J}}\xspace}
\newcommand{\Yhatd}{\ensuremath{\hat{Y}^{\dag}}\xspace}
\newcommand{\jhat}{\ensuremath{\hat{J}}\xspace}
\newcommand{\lhat}{\ensuremath{\hat{L}}\xspace}
\newcommand{\Nhat}{\ensuremath{\hat{N}}\xspace}
\newcommand{\rhohat}{\ensuremath{\hat{\rho}}\xspace}
\newcommand{\ddt}{\ensuremath{\frac{d}{dt}}\xspace}
\newcommand{\nset}{\ensuremath{n_1, n_2,\dots, n_k}\xspace}
\newcommand{\notes}[1]{{\color{blue}#1}}
\newcommand{\cmc}[1]{{\color{red}#1}}
\newcommand{\sah}[1]{{\color{magenta}#1}}

\title{Mean-field Dynamics and Fisher Information in Matterwave Interferometry}
\author{Simon A. Haine}
\email{simon.a.haine@gmail.com}
\affiliation{School of Mathematics and Physics, University of Queensland, Brisbane, QLD, 4072, Australia}

\begin{abstract}
There has been considerable recent interest in the mean-field dynamics of various atom-interferometry schemes designed for precision sensing. In the field of quantum metrology, the standard tools for evaluating metrological sensitivity are the classical and quantum Fisher information. In this letter, we show how these tools can be adapted to evaluate the sensitivity when the behaviour is dominated by mean-field dynamics. As an example, we compare the behaviour of four recent theoretical proposals for gyroscopes based on matterwaves interference in toroidally trapped geometries. We show that while the quantum Fisher information increases at different rates for the various schemes considered, in all cases it is consistent with the well-known Sagnac phase shift after the matterwaves have traversed a closed path. However, we argue that the relevant metric for quantifying interferometric sensitivity is the classical Fisher information, which can vary considerably between the schemes. 
\end{abstract}

\maketitle

\emph{Introduction---} Quantum devices based on matterwave interferometry, such as atom interferometers \cite{Cronin:2009}, atomic Josephson Junctions \cite{Ryu:2013} and Superfluid Helium Quantum Interference Devices (SHeQuIDS) \cite{Varoquax:2015} have the potential to provide extremely sensitive measurements of inertial quantities such as rotations, accelerations, and gravitational fields \cite{Riehle:1991, Gustavson:1997, Lenef:1997, Gustavson:2000, Durfee:2006, Canuel:2006,  Wu:2007, Stockton:2011, Dickerson:2013}. While the principles of matterwave interferometers are well understood, in practice, characterising and optimising interferometry schemes is still challenging, as there are many competing effects that can affect the sensitivity \cite{Szigeti:2012, Hardman:2014, McDonald:2014}. 

While there have recently been proof-of-principle demonstrations of matterwave interferometers displaying non-trivial quantum correlations \cite{Hald:1999, Kuzmich:2000, Gross:2010, Riedel:2010, Leroux:2010, Lucke:2011, Chen:2011, Sewell:2012, Hamley:2012}, to date, all matterwave interferometers with inertial sensing capabilities have been well described by \emph{mean-field} dynamics, which can be obtained by solving either the single particle Schr\"{o}dinger equation, or the Gross-Pitaevskii equation (GPE) \cite{Dalfovo:1999}. For example, there have been several recent proposals for atomic gyroscopes based on interference of Bose condensed atoms (BECs) confined in toroidal geometries, or `ring traps' \cite{Halkyard:2010, Kandes:2013, Helm:2015, Stevenson:2015, Nolan:2015, Bell:2015}. The analysis of these schemes has largely been concerned with the complex multi-mode dynamics of the \emph{order-parameter} $\psi(\boldr,t)$, which displays rich mean-field dynamics due to the inter-atomic interactions.  

The field of quantum metrology has developed sophisticated tools for evaluating the sensitivity of measurement devices, such as the quantum Fisher information (QFI) and the classical Fisher information (CFI) \cite{Toth:2014}. However, such analyses are usually concerned with the development of optimal measurement strategies with exotic quantum states, with the goal of providing measurement sensitivities better than the Standard Quantum Limit (SQL) \cite{Wineland:1992}, and largely ignore the classical effects that dominate matterwave interferometry, such as maximising interrogation times and mode-matching, with which mean-field analyses are concerned. In this letter, we demonstrate how to calculate the QFI and CFI from the mean-field dynamics of the system, and demonstrate that this is a useful method of quantifying the sensitivity even in the absence of quantum correlations. We apply this technique to four recently proposed schemes \cite{Halkyard:2010, Kandes:2013, Helm:2015, Stevenson:2015} concerning matterwave interferometry in ring traps, and show that this technique is very effective at identifying the advantages and disadvantages of each scheme.

\emph{Mean-Field Dynamics and Fisher Information---} The fundamental question when assessing the sensitivity of a matterwave interferometer is: By making measurements on the distribution of particles that have been effected by some classical parameter $\chi$ (which may be, for example, a parameter quantifying the magnitude of a rotation, acceleration, or gravitational field), how precisely can $\chi$ be estimated? The answer is given by the Quantum Cramer-Rao Bound (QCRB) \cite{Braunstein:1994}, which dictates that the smallest resolvable change in $\chi$ is $\delta \chi = \frac{1}{\sqrt{\mathcal{F}_Q}}$ where $\mathcal{F}_Q$ is the \emph{quantum Fisher information} (QFI), which for pure-states is $\mathcal{F}_Q = 4[\langle \dot{\Psi} | \dot{\Psi}\rangle - |\langle \Psi | \dot{\Psi}\rangle |^2]$, where $|\dot{\Psi}\rangle = \frac{\partial}{\partial \chi} |\Psi\rangle$ \cite{Toth:2014, Demkowicz-Dobrzanski:2014}. The analysis in \cite{Halkyard:2010, Kandes:2013, Helm:2015, Stevenson:2015} are largely concerned with the complicated multi-mode \emph{mean-field} dynamics of the \emph{order-parameter} $\psi(\boldr,t)$, which is simulated via the GPE \cite{Dalfovo:1999}, from which the mean density distribution can be calculated. The QFI is not normally considered in a mean-field analysis, as these calculations are agnostic about the form of the full quantum state $|\Psi\rangle$. While the order parameter $\psi(\boldr)$ is not usually considered as a quantum object, by assuming that the full $N$-particle state of the system is uncorrelated, we can use $\psi(\boldr)$ to calculate the QFI. Specifically, we make the reasonable assumption \cite{Leggett:2001} that $|\Psi(t)\rangle  = ((\hat{a}^\dag_\psi(t))^N / \sqrt{N!}) |0\rangle$, where $\hat{a}_\psi(t) = \int_{\mathbf{R}^3} \psi^*(\boldr,t) \hat{\psi}(\boldr) \, \dr$, 
or equivalently, that the system is represented by a many-body wavefunction of the form $\Psi(\boldr_1, \boldr_2, \dots, \boldr_N) = \psi(\boldr_1)\psi(\boldr_2)\dots\psi(\boldr_N)$. Due to the additive nature of QFI for separable systems \cite{Demkowicz-Dobrzanski:2014}, the QFI becomes $\mathcal{F}_Q = N F_Q$, where
\begin{equation}
F_Q = 4\left[ \int_{\mathbf{R}^3} \dot{\psi^*}\dot{\psi} \dr - \abs{ \int_{\mathbf{R}^3} \psi^*\dot{\psi} \dr}^2 \right] \label{F_Q_def}
\end{equation} 
is the \emph{single particle} QFI, and $\dot{\psi} = \frac{\partial}{\partial \chi} \psi(\boldr,t)$. The QFI tells us in principle how much information about the parameter $\chi$ that the state $|\Psi\rangle$ contains, assuming that we have complete freedom in the choice of measurement. However, in the case of matterwave interferometry, we are usually limited to making measurements of the spatial distribution of particles, as is the case via optical fluorescence, absorption, or phase-contrast imaging \cite{Ketterle:1999}, or detection via mulit-channel arrays such as is common in experiments with meta-stable Helium \cite{Vassen:2012}. Due to the nature of these imaging techniques, only two-dimensions of the spatial distribution at a single snapshot in time can be obtained, with the third dimension integrated over \cite{Ketterle:1999}. In this case we are restricted to the information that is contained in spatial probability distribution, and the sensitivity is limited to $\Delta \chi = \frac{1}{\sqrt{\mathcal{F}_C}}$, where $\mathcal{F}_C$ is the \emph{classical} Fisher information (CFI). Again, assuming that our many-body quantum state is uncorrelated, we can view the detection of the position of each atom as $N$ uncorrelated events, such that the CFI is simply $\mathcal{F}_C = N F_C$, where 
\begin{equation}
F_C = \int_{\mathbf{R}^2} \frac{1}{P(x,y)} \left( \frac{\partial P}{\partial \chi }\right)^2 dx dy \, , \label{F_C_def}
\end{equation}
and $P(x,y) = \int |\psi(\boldr,t)|^2 dz$, where we have chosen the $z$ direction as the imaging axis. The CFI quantifies how precisely we can estimate $\chi$ based purely on measurements of the two dimensional position distribution function. By optimising over all possible measurements it can be shown that $\mathcal{F}_Q \geq \mathcal{F}_C$ \cite{Toth:2014}. 

Obviously by assuming that our state is uncorrelated, as with all mean-field treatments,  we are ignoring the effects of any possible quantum correlations between the particles. However, in all matterwave interferometer inertial sensors so far demonstrated, the atomic sources are well approximated by uncorrelated systems \cite{Cronin:2009}. Additionally, in many of these experiments, the detection efficiency is low, or there are significant sources of loss \cite{Szigeti:2012} which has the effect of diminishing the importance of any correlations. 

\begin{figure*}
\centering
\includegraphics[width = 1\textwidth]{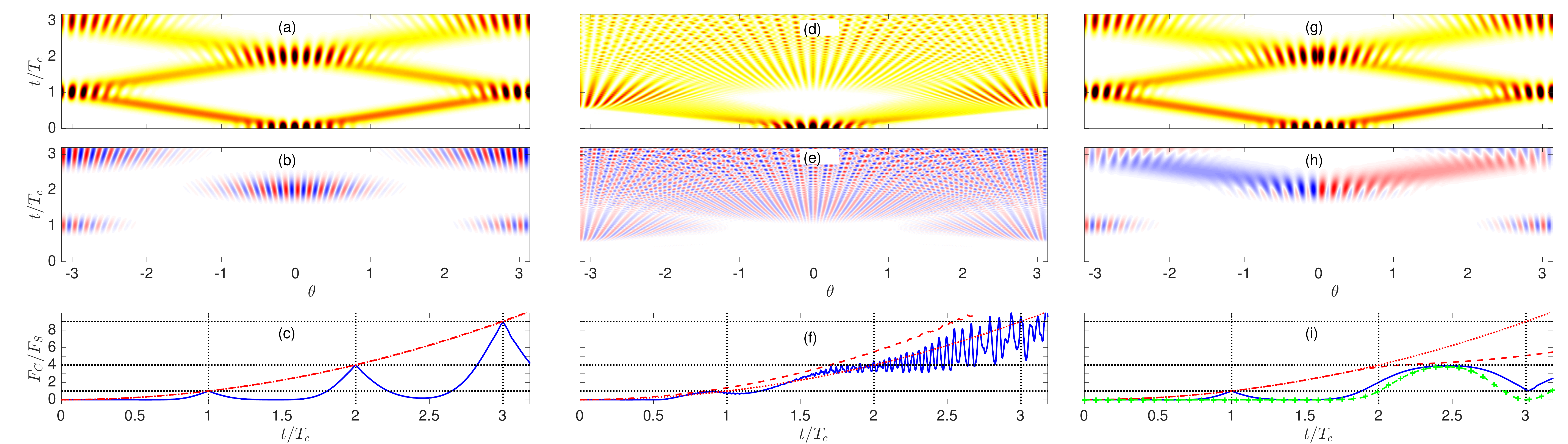}
\caption{\label{bromfig1} (Color Online) (a), (d), (g): $P(\theta, t)$ simulated with $\Omega = 0$ (Low density: white. High density: dark). (b), (e), (h): $d P(\theta,t)/d\Omega \vert_{\Omega =0}$ (Positive: red. Negative: blue. Zero: white). (c), (f), (i): $F_Q$ (red dashed line), $F_C$ (blue solid line), $F_{LR}$ (green $+$ symbols, (i) only), and $4t^2R^2 k_\textrm{kick}$ (red dotted line). The vertical dotted lines indicate integer multiples of the classical collision time, $T_c$, and the horizontal dotted lines indicate $n^2 F_S$, for integer values of $n$, indicating the number of closed loops the wavepackets have traversed. Parameters: For all frames: $\psi(\theta, 0) = \sqrt{2}(\sigma \sqrt{\pi/2})^{-\frac{1}{2}} \exp (-\theta^2/2\sigma^2) \cos (k_\textrm{kick} R \theta)$, $\sigma = 0.5$ rad,  $k_\textrm{kick} = 20/R$. (a)-(c) and (g)-(i): $U=0$, (d)-(f): $U = 0.2\hbar/mR$. In frames (g)-(i), a repulsive delta-function potential was introduced at $\theta=0$ and $t=T_c$.}
\end{figure*}

\emph{Comparison of Matterwave Gyroscopes---} When a matterwave in a rotating frame is split such that it traverses two separate paths enclosing an area $A$, the components in each path accumulate a phase difference given by the well-known Sagnac effect 
\begin{equation}
\phi_S = \frac{2m}{\hbar} \mathbf{\Omega}\cdot \mathbf{A}  \, , \label{sag_phase}
\end{equation}
where $m$ is the mass of the particle, $\mathbf{A} = A\hat{\mathbf{n}}$, where $\hat{\mathbf{n}}$ is the unit vector normal to the enclosed area, and $\mathbf{\Omega}$ is the angular velocity \cite{Cronin:2009}. We now turn our attention to the specific case of an interferometric matter wave gyroscope confined in a ring trap. In particular, we aim to use $F_Q$ and $F_C$ as a tool to compare the recent theoretical proposals \cite{Halkyard:2010, Kandes:2013, Helm:2015, Stevenson:2015}. Our aim is not to replicate every detail of these proposals, but to demonstrate how $F_C$ and $F_Q$ illuminate important aspects and the advantages and disadvantages of each scheme.  As in \cite{Halkyard:2010, Helm:2015, Kandes:2013}, working in cylindrical coordinates $\{r, z, \theta\}$, we assume a trapping potential of the form $V(\boldr) = \frac{1}{2} m\left(\omega_z^2 z^2 + \omega_r^2 (r-R)^2\right)$, where $R$ is the radius of the torus and $\omega_r$ and $\omega_z$ are the radial and axial trapping frequencies. Assuming that the radial and axial confinement is sufficiently tight, we may ignore the dynamics in these directions, in which case the evolution of the order parameter is described by the equation
\begin{equation}
i\hbar \frac{d}{dt} \psi(\theta, t)= \left(\frac{-\hbar^2}{2mR^2}\frac{\partial^2}{\partial \theta^2} + U|\psi|^2 -\Omega \hat{L}_z \right)\psi(\theta,t) \, , \label{GPE}
\end{equation}
where $\hat{L}_z$ is the $z$ component of the angular momentum, and we have assumed that we are working in a frame rotating around the $z$ axis at angular frequency $\Omega$. The goal of the device is to estimate $\Omega$ based on measurements of the matterwaves. We first restrict ourself to the non-interacting case $U=0$. In this case, $\hat{L}_z$ commutes with the other terms in the Hamiltonian which allows us to solve for the dynamics of $\psi(\theta,t)$ analytically: $\psi(\theta, t) = \hat{U}_\Omega \hat{U}_{KE}\psi(\theta,0)$, where $\hat{U}_\Omega = \exp \left(i \Omega \hat{L}_z t/\hbar\right)$, and $\hat{U}_{KE} = \exp \left(it \hbar/2m R^2 \frac{\partial^2}{\partial \theta^2} \right)$, which allows us to evaluate 
\begin{equation}
F_Q(t) = 4 t^2 V( \hat{L}_z /\hbar) \, ,
\end{equation}
where the variance may be computed with respect to either the initial state $\psi(\theta, 0)$ or the state at some later time $\psi(\theta,t)$. From this we see that initial states with a large spread in angular momentum will accumulate QFI more rapidly. To evaluate $F_C$, we  solve for $\psi(\theta,t)$, calculate $P(\theta,t) = |\psi(\theta,t)|^2$ for a range of different values of $\Omega$, and then calculate the derivative in \eq{F_C_def} numerically. We first examine the scheme proposed by Kandes \etal \cite{Kandes:2013}. They simulate a gaussian wavepacket (centred at $\theta =0$, initially at rest in the rotating frame), which is then split into two counter-propating components with momentum $\pm \hbar k_\textrm{kick}$. The wavepackets then traverse the ring in oposite directions, colliding (and passing through each other) on the far side of the ring ($\theta = \pi$) , and again back at $\theta = 0$. Fig.~(\ref{bromfig1}a) shows $P(\theta,t)$, which displays high-contrast interference fringes as the wavepackets pass through one another. The position of these fringes depends on the value of $\Omega$ used in the simulation. Fig.~(\ref{bromfig1}b) shows $d P(\theta,t)/d\Omega$, generated by performing simulations with slightly different values of $\Omega$. It can be seen that the derivative is negligible except when the wavepackets are overlapping. The asymmetric nature of the derivative indicates that small deviations in $\Omega$ can be inferred from the spatial position of the fringes.  Fig.(\ref{bromfig1}c) shows $F_Q$ and $F_C$ vs. time. As expected, $F_Q$ displays quadratic time-dependence with pre-factor $V(\hat{L}_z/\hbar) \approx R^2 k_\textrm{kick}^2$. The CFI is initially zero, but when the wavepackets begin to overlap, $F_C$ increases such that $F_C \approx F_Q$. The times at which this occurs is at integer multiples of the classical collision time $T_c = \pi R m/\hbar k_\textrm{kick}$, at which $F_Q \approx F_S =  (2 m \pi R^2 /\hbar)^2$, where $F_S$ is defined as the QFI of a state where the phase of two components differs by the Sagnac phase shift: $|\Psi\rangle = \frac{1}{\sqrt{2}}(|\psi_1\rangle + |\psi_2\rangle e^{i \phi_S})$, where $\phi_S$ is given by \eq{sag_phase}, and $\langle \psi_i |\psi_j\rangle = \delta_{ij}$. This quantity represents idealised operation of a matter-wave gyroscope after one closed loop has been traversed. From this analysis, we see that the magnitude of $k_\textrm{kick}$ increase the rate at which $F_Q$ accumulates, but it ultimately doesn't affect the value of $F_Q$ after an integer number of closed loops have been traversed. As we are restricted to measurements of the spatial distribution of particles, $F_C$ is the relevant quantity, which is sharply peaked around integer multiples of $T_c$, indicating that it is crucial to make the measurement at the collision times. 


So far these result are not particularly surprising. However, this analysis allows us to deal with more complicated systems where our analytic insight breaks down. One such example is by including a nonlinear interaction $U\neq0$ in \eq{GPE}. Fig.~(\ref{bromfig1}d) shows an identical simulation to fig.~(\ref{bromfig1}a), except with $U = 0.2 \hbar/mR$. The wavepackets now disperse much more rapidly until they become larger than the circumference of the ring, and the notion of a classical collision time and Sagnac phase shift becomes ill-defined. However, our Fisher information analysis sheds some light on the usefulness of this device (fig.(\ref{bromfig1})f). $F_Q$ increases more rapidly than the non-interacting case, and $F_C$ is no longer sharply peaked around integer multiples of $T_c$. Although $F_C$ is less than $F_Q$ for all time, it is also significantly greater than zero, and can be greater than $F_S$, indicating the existence of a method of processing the information contained in $P(\theta)$ in order to extract $\Omega$, even when the concept of the Sagnac phase shift \eq{sag_phase} become irrelevant due to different momentum components traversing different number of closed loops. Kandes \etal \cite{Kandes:2013} provide a method of extracting the phase shift based on analysing different frequency components of the density distribution, but this method assumes perfect signal-to-noise ratio and can not make predictions on the metrological sensitivity of the device, which our analysis does. 

In both of the above calculations, the rotational information is contained in the position of the interferences fringes in the density. This would require high-resolution spatial imagining, which could be challenging if the wavelength of the fringes becomes small. Helm \etal \cite{Helm:2015} model a similar  scheme, except that each wavepacket partially reflects off a sharp delta-function `barrier' at $\theta=0$, acting as a matterwave beamsplitter to convert the phase information into population information of the two counter-propagating wavepackets. The height of the barrier is tuned such that the wavepackets undergo 50\% quantum reflection, and the clock-wise and counter-clockwise propagating components can interfere. Fig.~(\ref{bromfig1}g) shows that the system behaves identically to that of Kandes \etal until the wavepackets encounter the barrier at $t\approx 2T_c$, after which time the relative populations of the counter-propagating wavepackets depends on $\Omega$. This is reflected in $F_C$, which displays a plateau of $F_C \approx 4 F_S$ after $2T_c$ until the wavepackets collide again, creating ambiguity in the population of each wavepacket. If our imaging system cannot fully resolve the details of the density distribution, but can distinguish between the right-going and left-going matterwave components, then the appropriate CFI is $F_{LR} = \sum_{j=L,R} P_j^{-1} \left(\frac{\partial P_j}{\partial \Omega}\right)^2$, where $P_L = \int_{-\pi}^0 P(\theta) d\theta$, $P_R = \int_{0}^\pi P(\theta) d\theta$ are the components of the matterwave on the left and right of the barrier respectively. Fig.~(\ref{bromfig1}i) shows that $F_{LR}$ is comparable to $F_C$, indicating that a measurement of the fraction of atoms on either side of the barrier is sufficient to extract the rotation information from the system. We note that although Helm \etal focus on the soliton regime for their simulations, we see that by simply using non-interacting wavepackets, $F_{LR}$ approaches $F_S$, indicating that this approach is sufficient to observe the full information from the Sagnac effect, without the need for operating in the soliton regime. 

Halkyard \etal \cite{Halkyard:2010} consider a different approach, where the matterwave is initially in the ground state of the potential, which uniformly fills the ring. A coupling pulse is then used to coherently transfer 50\% of the population to a different spin state while also transferring orbital angular momentum $\hbar \ell$ to this component. The two components remain spatially overlapped but accumulate a phase difference at a rate $\Delta \phi = 2\ell \Omega t$, which is then converted into either a population difference or density modulation between the two components via Ramsey interferometry. For simplicity, and as it highlights the important features of the scheme, we will initially consider only a single spin state, consisting of an equal superposition of $\hat{L}_z$ eigenstates with eigenvalues $\pm \hbar \ell$: $\psi(\theta, 0) = \frac{1}{\sqrt{4 \pi}}(e^{i\ell \theta} + e^{-i\ell \theta})$. In this case we have an exact expression for the variance of $\hat{L}_z$: $V(\hat{L}_z) = \hbar^2 \ell^2$, and $F_Q = 4 \ell^2 t^2$. Furthermore, it is trivial to solve for $\psi(\theta,t)$, which allows us to calculate the probability distribution $P(\theta, t) = 1 + \cos \left( 2\ell (\theta + \Omega t)\right)$, from which we can calculate the $F_C =4 \ell^2 t^2 =F_Q$, indicating that a measurement of the density saturates the QCRB for all time. That is, as the wavepackets are spatially overlapping for all times, information about the phase due to angular rotation can be observed in the density as persistent interference fringes. 

A common technique for overcoming the requirement for high spatial resolution is to use an additional degree of freedom such as the atomic spin \cite{Borde:1989}. If our two spin states are $|+1\rangle$ and $|-1\rangle$, then a general single particle state is $|\psi\rangle = \psi_{+1}(\boldr)|+1\rangle + \psi_{-1}(\boldr)|-1\rangle$. If our $N$-particle state is simply an uncorrelated product state, then $\mathcal{F}_Q = NF_Q$ where $F_Q = 4(\langle \dot{\psi} | \dot{\psi}\rangle - |\langle \dot{\psi}|\psi\rangle|^2)$.  By coherently coupling these two spin states via either a microwave or Raman transition, the phase information can be converted into population information, such that a measurement of the total number of particles in spin state, rather than the spatial distribution, is all that is required. If we restrict ourselves to measurements of the population of each spin state, then $F_C = \sum_{j=+1,-1} P_j^{-1}(d P_j /d\Omega)^2$, where $P_j = |\langle j|\psi\rangle|^2$ is the probability of finding each particle in the spin state $|j\rangle$. We now return to the example of Halkyard \etal, who prepare an initial state such that $\psi_{\pm 1}(\theta,0) = \frac{1}{2\sqrt{\pi}}e^{\pm i\ell \theta}$, which after time $T$ evolves to  $\psi_{\pm 1}(\theta,T) = \frac{1}{2\sqrt{\pi}}e^{\pm i\ell \theta}e^{\pm i\ell\Omega T}$. The two spin components are then coupled via a coherent Raman transition which transfers $2\ell$ units of orbital angular momentum, such that at the final time $t_f$ the state is $\psi_{\pm1}(\theta, t_f) = \frac{1}{\sqrt{2}}\left(\psi_{\pm 1}(\theta,T) -i\psi_{\mp 1}(\theta, T) e^{\pm 2i\ell \theta}\right)$. From this expression its simple to calculate the Fisher information and arrive at $F_Q = F_C = 4 \ell^2 T^2$. 

\begin{figure}
\includegraphics[width=1.0\columnwidth]{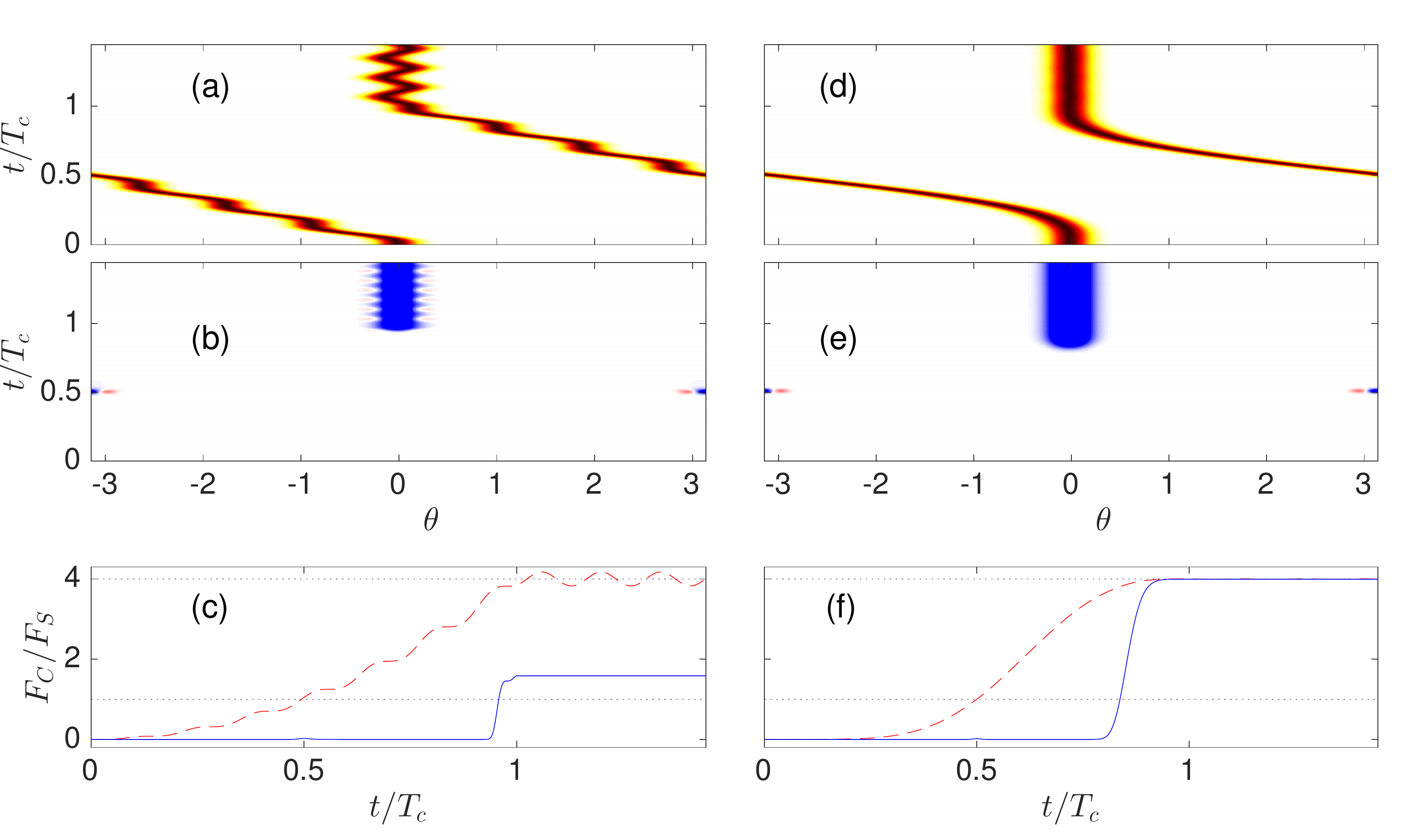}
\caption{ \label{clockfig}(Color Online) (a) \& (d): $|\psi_{+1}(\theta,t)|^2$ (Low density: white. High density: dark). $|\psi_{-1}(\theta, t)|^2$ is identical to $|\psi_{+1}(\theta, t)|^2$ except reflected around $\theta=0$. (b) \& (e): $\partial J_z(\theta, t)/\partial\Omega$, where $J_z =\frac{1}{2}( |\psi_{+1}(\theta,t_f)|^2 - |\psi_{-1}(\theta,t_f)|^2)$ (Positive: red. Negative: blue. Zero: white). (c) \& (f): $F_Q$ (red dashed line), $F_C$ (blue solid line). Each component was initially in the ground state of a spin dependent, harmonic trapping potential $V_\pm(\theta) =\frac{1}{2}m\omega_\theta^2 R^2(\theta-\theta_0)^2$. In (a), (b), and (c), the trap minimum moved with constant velocity: $\theta_0(t) = \mp 2\pi t/T_c$, and in (d), (e), (f), the trap minimum moved with a sinusoidal velocity profile: $\theta_0 = 2 \pi t/T_c - \sin (2 \pi t/T_c)$. Parameters: $R=5\sqrt{\hbar/m\omega_\theta}$, $T_c = 5\omega_\theta^{-1}$. }
\label{scheme}
\end{figure}

Finally, we consider the case of Stevenson \etal \cite{Stevenson:2015}, who depart from the notion of freely propagating matterwaves, and consider  two spin components $|+1\rangle$ and $|-1\rangle$, where the trapping potential for each component can be manipulated independently. The two spin components are transported around a closed loop in opposite directions via a time-dependent trapping potential, and then recombined via a microwave coupling pulse at time $T$ such that the state of the system at the final time $t_f$ is $\psi_{\pm1}(\theta, t_f) = \frac{1}{\sqrt{2}}\left(\psi_{\pm 1}(\theta,T) -i\psi_{\mp 1}(\theta, T) \right)$. Fig.~(\ref{clockfig}) shows the density distribution for one component, $\partial J_z(\theta, t)/\partial\Omega$, where $J_z =\frac{1}{2}( |\psi_{+1}(\theta,t_f)|^2 - |\psi_{-1}(\theta,t_f)|^2)$, and $F_C$ and $F_Q$ for two different cases. In the first case, the minimum of the harmonic trapping potential for each component moves from $\theta = 0$ to $\theta = \pi$ with constant velocity, which creates a centre of mass ``sloshing'' excitation, which inhibits the overlap of the two components such that $F_C$ is significantly less than $F_Q$. In the second case, the potential minimum moves with a sinusoidal velocity profile which creates far less mechanical excitation, and $F_C \approx F_Q$. 

\emph{Conclusion---} We have shown that both the CFI and QFI are useful tools for evaluating the mean-field dynamical aspects of matterwave interferometry. The quantum Fisher information automatically accounts for any phase information, even in cases where a simple notion of a phase shift may be ill-defined, or when there is no simple analytic expression for the phase evolution. The CFI automatically accounts for any issues of imperfect wave-packet overlap, and is the appropriate metric for the metrological information that can be extracted from measurements of the density distribution. This theoretical technique may be useful for analysing the sensitivity of devices where the dynamics is dominated by mean-field effects, such as atomic Josepheson junctions, or SHeQuIDs. 

The author would like to acknowledge useful discussions with Michael Bromley, Robin Stevenson, Sam Nolan, Stuart Szigeti, and Matthew Davis. The numerical simulations were performed with XMDS2 \cite{Dennis:2013}. This work was supported by Australian Research Council (ARC) Discovery Project No. DE130100575. 

\bibliography{simon_bib}

\end{document}